# Understanding and tuning magnetism in layered Ising-type antiferromagnet FePSe$_3$ for potential 2D magnet

*Rabindra Basnet\*, Taksh Patel, Jian Wang, Dinesh Upreti, Santosh Karki Chhetri, Gokul Acharya, Md Rafique Un Nabi, Josh Sakon, and Jin Hu\**

R Basnet, D Upreti, S. K. Chhetri, G Acharya, M. R. U. Nabi, J Hu
Department of Physics
University of Arkansas
Fayetteville, Arkansas 72701, USA
E-mail: basnetr@uapb.edu; jinhu@uark.edu

M. R. U. Nabi, J Hu
MonArk NSF Quantum Foundry
University of Arkansas
Fayetteville, Arkansas 72701, USA

R Basnet
Department of Chemistry & Physics
University of Arkansas at Pine Bluff
Pine Bluff, Arkansas 71603, USA

J Hu
Materials Science and Engineering Program
Institute for Nanoscience and Engineering
University of Arkansas
Fayetteville, Arkansas 72701, USA

T Patel
Fayetteville High School
Fayetteville, Arkansas 72701, USA

J Wang
Department of Chemistry and Biochemistry
Wichita State University
Wichita, KS 67260, USA

J Sakon
Department of Chemistry & Biochemistry
University of Arkansas
Fayetteville, Arkansas 72701, USA

**Abstract**

Recent development in two-dimensional (2D) magnetic materials have motivated the search for new van der Waals magnetic materials, especially Ising-type magnets with strong magnetic anisotropy. Fe-based *M*P*X*$_3$ (*M* = transition metal, *X* = chalcogen) compounds such as FePS$_3$ and FePSe$_3$ both exhibit an Ising-type magnetic order, but FePSe$_3$ receives much less attention compared to FePS$_3$. This work focuses on establishing the strategy to engineer magnetic anisotropy and exchange interactions in this less-explored compound. Through chalcogen and metal substitutions, the magnetic anisotropy is found to be immune against S substitution for Se whereas tunable only with heavy Mn substitution for Fe. In particular, Mn substitution leads to a continuous rotation of magnetic moments from the out-of-plane direction towards in-plane. Furthermore, the magnetic ordering temperature displays non-monotonic doping dependence for both chalcogen and metal substitutions but due to different mechanisms. These findings provide deeper insight into the Ising-type magnetism in this important van der Waals material, shedding light on the study of other Ising-type magnetic systems as well as discovering novel 2D magnets for potential applications in spintronics.

## 1. Introduction

The study of two-dimensional (2D) magnetic materials has greatly advanced our understanding of magnetism in low dimensions and the implementation of materials for technological applications[1–24]. So far, the studies have been limited to a few material systems. Seeking new magnetic van der Waals (vdW) materials with potential to realize 2D magnetism and engineering their magnetic properties has become one important research direction. With this motivation, vdW-type antiferromagnetic (AFM) $MPX_3$ ($M$ = transition metal, $X$ = chalcogen) materials have attracted growing attentions owing to their well-established magnetic orders in bulk materials and the feasibility of obtaining their atomically thin layers[5,21–43]. Importantly, magnetism in $MPX_3$ varies with the choice of $M$ and $X$[25–42], which has motivated numerous efforts to tune magnetic properties such as substitutions for metal $M$[36,39,41,44–56] and chalcogen $X$[43,57–60], as well as inter-layer intercalation[61–63]. Such a tunable material platform offers rich opportunities for exploring 2D magnetism. So far, the study of 2D magnetism in exfoliated atomically thin $MPX_3$ flakes is still in the early stage, which has been limited to a few compounds such as $NiPS_3$[21], $MnPS_3$[5,23], $MnPSe_3$ [24], $FePS_3$[22], and $FePSe_3$[64]. Those studies have revealed that the persistence of magnetism in the 2D limit depends on the type of magnetic orders. Only compounds possessing strong magnetic anisotropy such as $MnPSe_3$[24], $FePS_3$[22], and $FePSe_3$[64] can maintain long-range magnetic order in their monolayer form. In 2D systems, it has been proposed that long-range magnetic orders are strongly suppressed by thermal fluctuations[65], which can be counteracted by magnetic anisotropy. Hence, the strength of magnetic anisotropy plays an important role in stabilizing magnetism in the 2D regime. Therefore, most of 2D magnets such as the atomically thin layers of $FePS_3$[22], $FePSe_3$[64], $CrI_3$[1], $CrBr_3$[66], $VI_3$[67] and $Fe_3GeTe_2$[3] display highly anisotropic Ising-type magnetism characterized by out-of-plane magnetic moments. Thus, studying Ising-type magnetic materials and further tuning their magnetism would provide insight into realizing 2D magnets with novel functionalities.

This work focuses on investigating the Ising-type antiferromagnet FePSe$_3$ through chalcogen and metal substitutions. We found that S and Mn substitutions in FePSe$_3$ play distinct roles in manipulating magnetic anisotropies and exchange interactions. Our work provides a better understanding of the Ising-type magnetism in FePSe$_3$ and related compounds, which can be further extended to other Ising-type systems. Furthermore, the realized tunable Ising-type magnetic material offer a novel platform to explore 2D magnetism and device applications.

## 2. Result and discussion

As a member of the *M*P*X*$_3$ family, FePSe$_3$ was discovered a few decades ago[43,68] but received surprisingly less attention than its sibling compound FePS$_3$[43,53,64,69]. To understand and tune the magnetism in FePSe$_3$, two substitution strategies, chalcogen and metal substitutions, have been adopted in this work. Chalcogen substitution, i.e., replacing S with Se or *vice versa*, has been found to be effective in modifying magnetic anisotropies in MnP(S,Se)$_3$ and NiP(S,Se)$_3$[59,70]. For FeP(S,Se)$_3$, the fully S-substituted compound FePS$_3$ has been identified as a representative *M*P*X*$_3$ material, which displays Ising-type magnetism characterized by out-of-plane magnetic moments (Figure 1a)[22,26,30]. Such Ising-type magnetism in FePS$_3$ has been proposed to stem from the strong spin-orbit coupling (SOC) of the high-spin Fe$^{2+}$ ($d^6$) state and the trigonal distortion of the FeS$_6$ octahedra[26]. Unlike many other *M*P*X*$_3$ compounds such as MnP(S,Se)$_3$ and NiP(S,Se)$_3$[59,70] which show distinct magnetic structures for sulfide (*M*PS$_3$) and selenide (*M*PSe$_3$), both FePS$_3$[22,26,30] and FePSe$_3$[43,64] exhibit similar Ising-type AFM ordering from bulk to the monolayer limit. This ordering is characterized by antiferromagnetically coupled FM zig-zag spin chains in each layer, as depicted in Figure 1(a). The presence of such a similar magnetic structure naturally raises the question of whether chalcogen substation may play a role in modifying magnetism, which will be addressed as shown below.

Metal substitution in $M$P$X_3$, unlike the chalcogen substitution which leaves the magnetic metal layer intact, introduces inevitable magnetic fluctuations and frustrations. Nevertheless, metal substitution has been demonstrated as a higly effective approach to control magnetism in $M$P$X_3$ due to the distinct single-ion anisotropy for different $M^{2+}$ ions[39,41,53,54,56]. FePSe$_3$ and MnPSe$_3$ studied in this work represent such examples. Given that the Fe moments are along the out-of-plane direction in FePSe$_3$, while the Mn moments mostly lie within the basal plane in MnPSe$_3$[24,43,57] (Figure 1a), elucidating the evolution of magnetism from the Fe side to the Mn side in Fe$_{1-x}$Mn$_x$PSe$_3$ would offer deep insights into the mechanism of magnetism in $M$P$X_3$ compounds and shed light on the control of magnetism.

As discussed above, the chalcogen S and metal Mn substitutions provide two distinct routes to control and further understand the magnetism in FePSe$_3$. However, the magnetic properties of S-substituted FePSe$_3$ have not been studied so far, and for metal substitution, only polycrystalline Fe$_{1-x}$Mn$_x$PSe$_3$ have been investigated[53]. This work focuses on single crystalline samples which can provide more insight into anisotropy, especially in magnetic property studies. Through extensive crystal growth efforts, we have obtained sizeable single crystals of FeP(Se$_{1-x}$S$_x$)$_3$ and Fe$_{1-x}$Mn$_x$PSe$_3$ ($0 \leq x \leq 1$). The successful S and Mn substitutions in FePSe$_3$ were demonstrated by composition analyses using energy-dispersive x-ray spectroscopy (EDS) and further confirmed by structure characterizations using x-ray diffraction (XRD). It has been reported that FePSe$_3$ shares a similar rhombohedra lattice structure with MnPSe$_3$[53] but is different from that of the monoclinic FePS$_3$ (space group $C2/m$). To examine the crystal structures of the substituted samples, we performed XRD experiments on powdered samples obtained by grinding single crystals. As shown in Figure 1b, the diffraction pattern for the pristine FePSe$_3$ can be well-indexed by the known rhombohedra structural model. In the case of S-substituted FeP(Se$_{1-x}$S$_x$)$_3$ samples (Figure 1b, upper panel), S substitution induces systematic high-angle peak shifts up to $x = 0.5$. Further increasing S content causes a structural crossover to the monoclinic FePS$_3$ type. It is worth noting that the $x = 0.66$ sample displays a more complicated XRD pattern, which has been found to be caused by the coexistence of both rhombohedra and monoclinic phases as confirmed by our Rietveld refinement. In addition, as shown in Figure 1b, this sample also displays an impurity peak that can be ascribed to the non-magnetic β-P$_4$S$_7$ phase which does not affect our property study. On the other hand, for Mn-substituted Fe$_{1-x}$Mn$_x$PSe$_3$ (Figure 1b, lower panel), metal substitution does not significantly alter the lattice structure but results in a systematic low-angle shift upon increasing Mn content, consistent with the lattice expansion due to the incorporation of larger Mn atoms.

To investigate the evolution of magnetic anisotropy in FeP(Se$_{1-x}$S$_x$)$_3$ and Fe$_{1-x}$Mn$_x$PSe$_3$, we have measured the temperature dependence of susceptibility ($\chi$) under out-of-plane ($H \perp ab$) and in-plane ($H // ab$) magnetic fields of $\mu_0 H = 0.1$ T. Because the sample holder may contribute to magnetic anisotropy[27], we have used the identical sample holder for both out-of-plane ($\chi_\perp$) and in-plane ($\chi_{//}$) susceptibility measurements. The contributions from the sample holder were separately measured and subtracted from the measured total magnetization data. As shown in Figure 2(a), the temperature dependencies for $\chi_\perp$ (solid line) and $\chi_{//}$ (dashed line) for chalcogen substituted FeP(Se$_{1-x}$S$_x$)$_3$ exhibit significant anisotropy both below and above the AFM transition temperature ($T_N$) (denoted by black triangles in Figure 2) for all sample compositions from $x = 0$ to 1. The anisotropic susceptibility above $T_N$ has been observed well beyond $T_N$ ($\approx$ 120 K) up to $T = 400$ K in pristine FePS$_3$[26,54,56]. In this work, we found such anisotropy extends to various Se-substituted FePS$_3$ and persists to fully Se-substituted compound FePSe$_3$. Such phenomena can be understood as follows: due to much weaker inter-layer interactions than in-plane interactions owing to the layered structure of $MPX_3$[30], these compounds are good approximation to 2D magnets. For such layered magnetic materials, a short-range 2D or quasi-2D magnetic correlation has been proposed to persist above $T_N$ in the paramagnetic (PM) phase[26], and this has been experimentally demonstrated by $^{31}$P nuclear magnetic resonance measurements[71]. This 2D- or quasi-2D magnetic correlation is reported to manifest as a broad maximum just above $T_N$ in temperature-dependent susceptibility for $M$P$X_3$[26], which has also been observed in our FeP(Se$_{1-x}$S$_x$)$_3$ samples as indicated by red triangles in Figure 2(a), suggesting the existence of short-range magnetic ordering in the PM phase of our FeP(Se$_{1-x}$S$_x$)$_3$ samples. Hence, the anisotropic susceptibility above $T_N$ in FeP(Se$_{1-x}$S$_x$)$_3$ might be related to these short-range magnetic correlations. It is worth noting that, though short-range magnetic correlations in the PM state should exists in all $M$P$X_3$ compounds, strong susceptibility anisotropy is not present in many other $M$P$X_3$ compounds such as MnP$X_3$[35,37,59] and NiP$X_3$[27,54,59,72]. This difference may be ascribed to the highly anisotropic Ising-type magnetism in FeP(Se$_{1-x}$S$_x$)$_3$, the magnetic correlation of which causes significant magnetic susceptibility anisotropy above $T_N$. The typical behavior for MnP$X_3$ and NiP$X_3$ have been attributed to their relatively weaker magnetic anisotropy[26,27,29]. Therefore, the observed strong anisotropy in FeP(Se$_{1-x}$S$_x$)$_3$ might be related to the Ising-type magnetic ordering in both FePSe$_3$[69] and FePS$_3$[22,26,56], suggesting the persistence of the Ising-type magnetic structure for the entire composition range.

In FeP(Se$_{1-x}$S$_x$)$_3$, the Ising-type magnetic structure upon substitution is supported by the unchanged magnetic easy axis. As shown in Figure 2(a), the susceptibilities for various FeP(Se$_{1-x}$S$_x$)$_3$ samples exhibit almost identical temperature dependence: $\chi_\perp$ displays drastically drop below $T_N$ while the variation of $\chi_{//}$ is much weaker, which is consistent with AFM ordering with an out-of-plane moment orientation. The unchanged magnetic easy axis against substitution in FeP(Se$_{1-x}$S$_x$)$_3$ is distinct from the switching of easy axis between in-plane and out-of-plane directions seen in Se-substituted MnPS$_3$ and NiPS$_3$[59]. Such difference is likely attributed to their different origins for magnetic anisotropy. The quenched or partially quenched orbital angular momentum for 3*d* transition metal ions leads to weak spin-orbit coupling (SOC) and consequently small single-ion anisotropy (*A*). In such a case, magnetic anisotropy mainly originates from anisotropic superexchange interactions that arises due to the SOC of non-magnetic ligands[73,74]. For example, the FM and AFM ground states in CrI$_3$[73] and MnPSe$_3$[74], respectively, are stabilized by ligands-mediated superexchange interactions. Hence, the modification of easy axis due to S-Se substitution is plausible in MnPS$_3$ and NiPS$_3$[58,70]. The situation is different for FePS$_3$ in which the strong crystal-field anisotropy of Fe$^{2+}$ ions[75,76] leads to a much higher *A* ($\approx$ 2.66 meV)[30] compared to MnPS$_3$ (*A* $\approx$ 0.0086 meV)[31] and NiPS$_3$ (*A* $\approx$ 0.3 meV)[77]. Therefore, the magnetic anisotropy in FePSe$_3$ and FePS$_3$ predominantly arises from the crystal-field anisotropy of Fe$^{2+}$ ions. Consequently, S-Se substitution has a less effect on the magnetic anisotropy in FeP(Se$_{1-x}$S$_x$)$_3$. Although the S substitution for Se in FePSe$_3$ leads to a crystal structure crossover from rhombohedra to monoclinic, the Ising-type magnetic ordering is robust.

Given that the magnetic anisotropy in FePSe$_3$ mainly originates from Fe$^{2+}$ crystal-field anisotropy, substitution in the Fe sites instead of Se should be a more effective way to tune anisotropy. This has indeed been demonstrated in our Mn-Fe metal substitution study. As shown in Figure 2(b), in contrast to the S-Se substitution which maintains the significant anisotropy between $\chi_\perp$ and $\chi_{//}$, the Mn substitution for Fe suppresses anisotropy above $T_\mathrm{N}$, as manifested by the overlapping of $\chi_\perp$ and $\chi_{//}$ in the PM state. This is suggestive of the variation of magnetic anisotropy with metal substitution, which eventually leads to the different magnetic structures for pristine FePSe$_3$ and MnPSe$_3$[43,53].

Tuning the magnetic anisotropy in *M*P*X*$_3$ corresponds to changing the magnetic easy axis[26,39,59]. The variation of the magnetic anisotropy in Mn-substituted FePSe$_3$ suggests a rotation of magnetic moments away from the out-of-plane direction of FePSe$_3$. However, Mn-substitution appears not very efficient in inducing such moment rotation. As mentioned above, the Ising-type AFM ordering in FePSe$_3$ leads to much stronger drop of $\chi_\perp$ than $\chi_{//}$ below $T_N$. Similarly, as shown in Figure 2(b), for a wide composition range from $x = 0$ to 0.9 in Fe$_{1-x}$Mn$_x$PSe$_3$, the much stronger drop of $\chi_\perp$ than $\chi_{//}$ below $T_N$ implies the easy axis is still along or close to the out-of-plane direction. The switching of anisotropy may occur in the $x = 0.93$ sample where $\chi_\perp$ slightly surpasses $\chi_{//}$ below $T_N$. Eventually, at $x = 1$, the pristine MnPSe$_3$ exhibits roughly constant $\chi_\perp$ but notably dropped $\chi_{//}$ in the AFM state, which is a typical behavior for an in-plane magnetic easy axis that has been verified by neutron scattering[43,57].

It is rather surprising that FePSe$_3$ maintains its magnetic anisotropy even with up to 90% of Mn substitution. Interestingly, a similar retention of anisotropy upon large Mn for Fe substitution (up to 97%) has also been observed in another Fe-based compound K$_2$FeF$_4$[76]. In addition to FePSe$_3$, the sulfide compound FePS$_3$ also exhibits a relatively rigid moment orientation. FePS$_3$[22,26,30] and NiPS$_3$[27] display distinct out-of-plane (Ising-type) and almost in-plane magnetic moment orientations, respectively. Previous studies have found that substituting 90% Ni for Fe is unable to modify the easy axis in FePS$_3$[54,75]. Therefore, in mixed systems that consists of two type of metal ions with different strength of single-ion anisotropies, a strongly anisotropic ion (like Fe$^{2+}$) dictates the ion with weaker anisotropy (like Mn$^{2+}$ or Ni$^{2+}$) through exchange interaction[75]. Thus, the out-of-plane easy axis in FePSe$_3$ and FePS$_3$ remains robust against various metal substitutions up to 90%.

The spin rotation induced by heavy Mn substitution is also evident in the field-dependent magnetization measured under out-of-plane ($H\perp ab$) (red color) and in-plane ($H//ab$) (blue color) magnetic fields. As shown in Figure 3(b), the isothermal magnetization at $T = 2$ K displays linear field dependence up to $\mu_0H = 9$ T for $x = 0 - 0.36$ samples but exhibits a clear metamagnetic transition in $x = 0.79$ and $0.9$ samples (denoted by red arrows) under out-of-plane magnetic field. Such a metamagnetic transition has been observed in a few $MPX_3$ compounds and attributed to a spin-flop (SF) transition[38,39,41], which is characterized by the moment reorientation driven by the magnetic field component parallel to the magnetic easy axis. The linear field dependence for magnetization up to 9 T in pristine FePSe$_3$ ($x = 0$) is understandable, because its Ising-type magnetic ordering may require a strong magnetic field to drive moment reorientation. In fact, a high field study on sulfide sample FePS$_3$ has revealed that the magnetization transition occurs above $\mu_0H = 35$ T at $T = 4$ K[40]. As discussed earlier, the entire FeP(Se$_{1-x}$S$_x$)$_3$ family exhibits strong anisotropic magnetism, so linear field-dependent magnetization under both in-plane and out-of-plane fields up to $\mu_0H = 9$ T is not surprising [Figure 3(a)]. The scenario is different in Mn-substituted samples [Fig. 3(b)]. As mentioned above, substituting Mn for Fe pushes the easy axis towards the basal plane. This rotation of easy axis can suppress the SF field as seen in Ni-[39] and Se-substituted[59] MnPS$_3$. Therefore, the heavily Mn-substituted $x = 0.79$ and $0.9$ samples exhibit SF transitions under relatively lower out-of-plane ($H\perp ab$) magnetic fields. When the easy axis rotates towards the $ab$-plane in the $x = 0.93$ sample, the SF transition under $H\perp ab$ is absent but a weak metamagnetic transition appears for $H//ab$ (denoted by the blue arrow in the inset). Further increasing the Mn content to $x = 1$, a much clear metamagnetic transition appears at slightly lower in-plane field as indicated by the blue arrow in the inset of Figure 3(b), suggesting a possible SF transition in MnPSe$_3$ which is characterized by an in-plane easy axis [Figure 1(a)].

Our results demonstrate that Ising-type AFM ordering in FePSe$_3$ is unaffected by S substitution but can be tuned with havey Mn substitution. The strong anisotropy in FeP(Se$_{1-x}$S$_x$)$_3$ compounds makes them promising candidates for 2D magnets. In addition, given that both pristine FePSe$_3$[64] and MnPSe$_3$[24] exhibit 2D magnetism in the monolayer limit, the Mn-substituted FePSe$_3$ offers further opportunity for tuning 2D magnetism. Nevertheless, the strong frustration accompanied by Mn substitution, which arises from the mixing of two different magnetic metal ions, could destabilize magnetic order in the 2D limit. Frustration in metal-substituted $M$P$X_3$ compounds is evident in the evolution of the magnetic transition temperature ($T_N$). In polymetallic $M$P$X_3$ compounds[36,39,41,49], $T_N$ has been found to reduce with substitution until reaching a minimum value around $x = 0.5$ where frustration is maximized. To elucidate the impact of substitution on magnetism, we have summarized the composition dependence of magnetic transition temperatures for FeP(Se$_{1-x}$S$_x$)$_3$ and Fe$_{1-x}$Mn$_x$PSe$_3$. To obtain the precise transition temperature, we calculated the derivative d$\chi$/d$T$ for susceptibility data shown in Figure 2 and used their peak position to define $T_N$ [Figure 4(a)], which has been widely used in previous studies[41,54,56,59,72]. The extracted $T_N$ values for the end compounds FePSe$_3$, FePS$_3$, and MnPSe$_3$ are 111.1, 120.1, and 73.4 K, respectively, consistent with the reported values[28,30,43,53,59,64]. As shown in Figures 4(b) and (d), both Fe$_{1-x}$Mn$_x$PSe$_3$ and FeP(Se$_{1-x}$S$_x$)$_3$ samples exhibit similar non-monotonic composition dependent $T_N$.

For metal substituted Fe$_{1-x}$Mn$_x$PSe$_3$ compounds [Figure 4(b)], $T_N$ reaches a minimum at $x = 0.5$ following a scenario of magnetic frustration similar to the one discussed above, which has also been reported in the earlier polycrystal study[53]. As mentioned above, for Fe$_{1-x}$Mn$_x$PSe$_3$, the spin reorientation from the in-plane to the out-of-plane direction occurs at around $x = 0.9$, which is significantly different than the minimum $T_N$ at $x = 0.5$. The spin orientation and $T_N$ in $M$P$X_3$, though ultimately influenced by the competing effects introduced by two different metal ions, are primarily determined by different factors: The spin orientation is greatly affected by magnetic anisotropy, while the magnetic ordering temperature is determined by magnetic exchange interactions[26,39,59,77]. Substituting Mn for Fe in Fe$_{1-x}$Mn$_x$PSe$_3$ produces distinct effects on magnetic anisotropy and exchange, which may be estimated from the relative magnitudes of these parameters for the two end compounds FePSe$_3$ and MnPSe$_3$. However, their experimental values, though have been recently determined for MnPSe$_3$ by neutron scattering experiment[57], are still lacking for FePSe$_3$. This makes the direct comparison of each parameter for the two end compounds difficult. Fortunately, the sulfide counterparts FePS$_3$ and MnPS$_3$ can provide some insights. Neutron scattering experiments have revealed different but comparable magnetic exchange parameters (*J*) for these two compounds whereas the single-ion anisotropy (*A*) for FePS$_3$[30] is significantly higher (by more than 300 times) than that of MnPS$_3$[31]. Thus, in selendie samples, substitution of Fe for Mn may also affect magnetic anisotropy more efficiently than exchange interactions. This explains the sensitive tuning of spin orientation by only replacing 10% Mn by Fe in MnPSe$_3$.

For $T_N$, on the other hand, it is determined by magnetic exchange interactions within and between magnetic sublattices in AFM materials in a more complicated way. Therefore, though exchange parameters for the end compounds FePS$_3$ and MnPS$_3$ have comparable values[30][31], in mixed Fe$_{1-x}$Mn$_x$PSe$_3$, fluctuations and frustrations due to mixing two types of magnetic ions effectively supress exchange interactions. As a result, $T_N$ reaches a minimum at 50% substitution when fluctuations and frustrations are maximized. In the case of chalcogen substituted in FeP(Se$_{1-x}$S$_x$)$_3$, though chalcogen substitution does not directly modify the magnetic atom layers, we still observe a non-monotonic evolution for $T_N$, with a minimum value when half of Se is replaced by S [Figure 4(d)]. This behavior echoes a magnetic frustration scenario similar to that discussed above for metal substitution. It noteworthy that the suppression of $T_N$ in FePSe$_3$ induced by chalcogen substitution is much weaker than that caused by metal substitution. Specifically, $T_N$ is reduced by only 4.7% in case of 50% S substitution for Se, in contrast to a significant 67% reduction when half of Fe is replaced by Mn. Indeed, substituting ligands modifies only the local environment around metal atoms without affecting the magnetic layers. As a result, it is expected to induce much less frustrations compared to metal substitutions[59]. For example, previous studies on chalcogen substitutions in MnP(S,Se)$_3$ and NiP(S,Se)$_3$ have found distinct monotonic composition dependences for $T_N$, implying that chalcogen substitutions primarily tune magnetic interactions rather than inducing strong frustrations[59,60]. Thus, the non-monotonic dependence of $T_N$ in FeP(Se$_{1-x}$S$_x$)$_3$ might be relevant to the tuning of magnetic exchanges, as discussed below.

The overall magnetic interactions in MnPS$_3$ and NiPS$_3$ are governed by the nearest-neighbor ($J_1$) and the third nearest-neighbor ($J_3$) exchanges respectively[77]. In Se-substituted MnPS$_3$ and NiPS$_3$, the change in $T_N$ with substitutions have been ascribed to the systematic variation of the dominant $J_1$ and $J_3$, respectively[59,60]. In FePS$_3$, previous neutron scattering measurements have unveiled the dominant $J_1$[30], which is ferromagnetic (FM) in nature [Figure 4(c)] and sensitive to the Fe-Fe distance [77,78]. A mere 5% elongation of Fe-Fe distance in FePS$_3$ is found to substantially modify the nearest-neighbor Fe-Fe FM interaction[78]. Therefore, enhancing FM $J_1$ may consequently suppress the AFM ordering. Indeed, we observed a correlation between $T_N$ and the nearest-neighbor Fe-Fe distance. As shown in Figure 4(e), the nearest-neighbor Fe-Fe distance, obtained from Rietveld refinement of XRD patterns [Figure 1(b)], displays a non-monotonic dependence on composition. Initially, the Fe-Fe distance decreases with increasing S content up to $x = 0.5$, thereby enhancing FM $J_1$ and leading to a suppression of $T_N$, as illustrated in Figure 4(d). Subsequently, as the S content surpasses $x = 0.5$, the Fe-Fe distance elongates, which consequently enhances $T_N$ for these S-rich samples. Of course, the slight lattice changes from $R\bar{3}$ ($x \leq 0.5$) to $C2/m$ ($x > 0.5$) space group may also contribute to the modulation of $T_N$. Further theoretical studies are needed to better clarify the mechanisms behind the unusual non-monotonic evolution of $T_N$ in FeP(Se$_{1-x}$S$_x$)$_3$.

## 3. Conclusion

In conclusion, we have studied the magnetic properties of the Ising-type antiferromagnet FePSe$_3$ and identified strategies to engineer its magnetism. The magnetic anisotropy in pristine FePSe$_3$ is robust against S substitutions but more tunable with Mn substitutions. In addition, both S and Mn substitutions result in a non-monotonic evolution of the magnetic ordering temperature, which might be attributed to different mechanisms of Fe-Fe distance-mediated exchange interactions and magnetic frustrations, respectively. Our study provides a deeper understanding of the Ising-type Fe-based *MPX*$_3$ vdW magnetic system, offering an important platform for discovering novel 2D magnets and engineer magnetic properties.

## 4. Experimental Section

*Materials Synthesis*: The single crystals of FeP(Se$_{1-x}$S$_x$)$_3$ ($0 \leq x \leq 1$) and Fe$_{1-x}$Mn$_x$PSe$_3$ ($0 \leq x \leq 1$) used in this work were synthesized via a chemical vapor transport method using I$_2$ as the transport agent. For each composition, elemental powders with desired molar ratios were sealed in a quartz tube and placed in a two-zone furnace with a temperature gradient from 750 to 550 °C for a week.

*Elemental and structure characterizations*: The elemental compositions and crystal structures of the obtained crystals were examined by energy-dispersive x-ray spectroscopy (EDS) and x-ray diffraction (XRD), respectively.

*Magnetic property characterizations*: Magnetization measurements were performed in a physical property measurement system (PPMS, Quantum Design).

**Acknowledgements**

This work was primarily supported by the U.S. Department of Energy, Office of Science, Basic Energy Sciences program under Grant No. DE-SC0022006 (synthesis and magnetic property study). J. S. acknowledges the support from NIH under award P20GM103429 for powder XRD experiment. J.W. acknowledges the support from the U.S. National Science Foundation under grand DMR-2316811 for structure refinement and analysis.

**Conflict of Interest**

The authors declare no conflict of interest.

**Data Availability Statement**

The data that support the findings of this study are available from the corresponding author upon reasonable request.

**Keywords**

two-dimensional magnet, Ising-type magnetism, magnetic anisotropy, spin-flop transition

**Figure 1.**

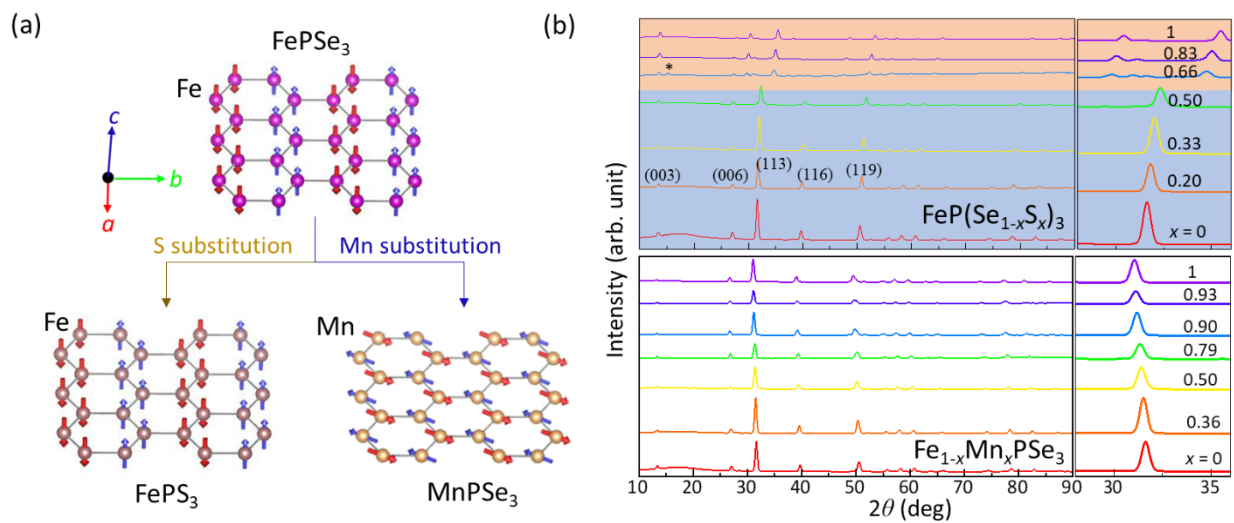

**Figure 1.** a) Magnetic structures of FePSe$_3$, FePS$_3$, and MnPSe$_3$. Only metal ions are shown. b) Powder x-ray diffraction patterns of FeP(Se$_{1-x}$S$_x$)$_3$ (upper panel) and Fe$_{1-x}$Mn$_x$PSe$_3$ (lower panel) (0 ≤ $x$ ≤ 1) samples. Different colored regions in the upper panel represent different crystal structure for FeP(Se$_{1-x}$S$_x$)$_3$ (orange: monoclinic, space group $C2/m$; blue: rhombohedra, space group $R\bar{3}$). The right panels show the evolution of the (113) diffraction peak of FePSe$_3$ with S and Mn substitutions. The Se and Mn contents for each sample are determined by EDS. The * for the $x$ = 0.66 sample of FeP(Se$_{1-x}$S$_x$)$_3$ marks the $\beta$-P$_4$S$_7$ impurity.

**Figure 2.**

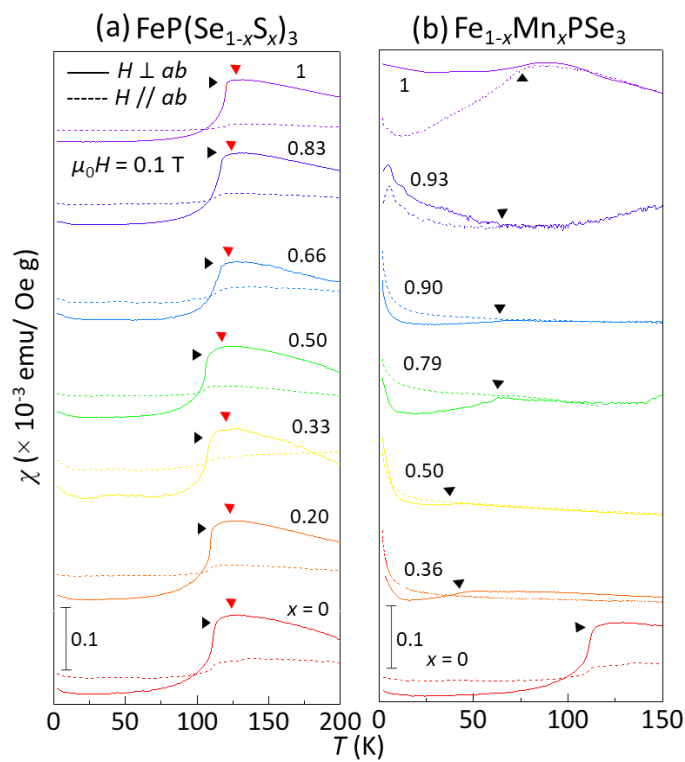

**Figure 2.** Temperature dependencies of the out-of-plane ($H\perp ab$, solid line) and the in-plane ($H\|ab$, dashed line) susceptibility ($\chi$) for a) FeP(Se$_{1-x}$S$_x$)$_3$ and b) Fe$_{1-x}$Mn$_x$PSe$_3$ ($0 \leq x \leq 1$) samples measured under magnetic field of 0.1 T. The black and red triangles denote $T_N$ and the susceptibility broad maximum, respectively. Data for different compositions are shifted for better comparison.

**Figure 3.**

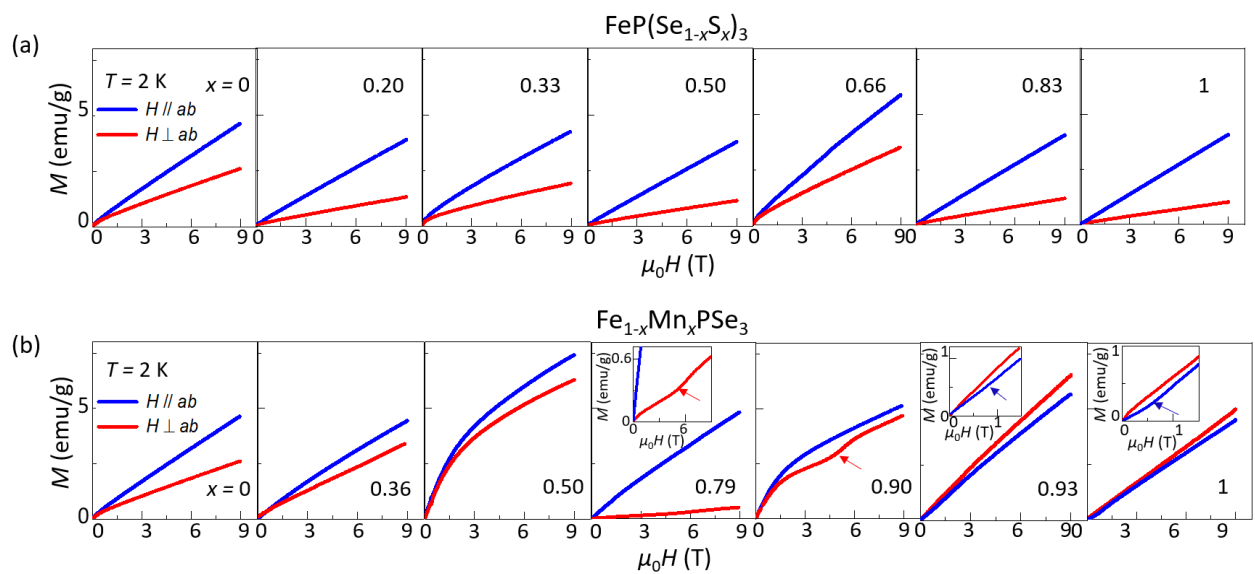

**Figure 3.** Isothermal magnetization at 2 K for a) FeP(Se$_{1-x}$S$_x$)$_3$ (0 ≤ $x$ ≤ 1) and b) Fe$_{1-x}$Mn$_x$PSe$_3$ (0 ≤ $x$ ≤ 1) samples measured under out-of-plane (*H*⊥*ab*, red) and in-plane (*H*∥*ab*, blue) magnetic fields. Inset: low-field magnetizations. The red and blue arrows in b) denote spin-flop fields under *H*⊥*ab* and *H*∥*ab* magnetic fields, respectively.

**Figure 4.**

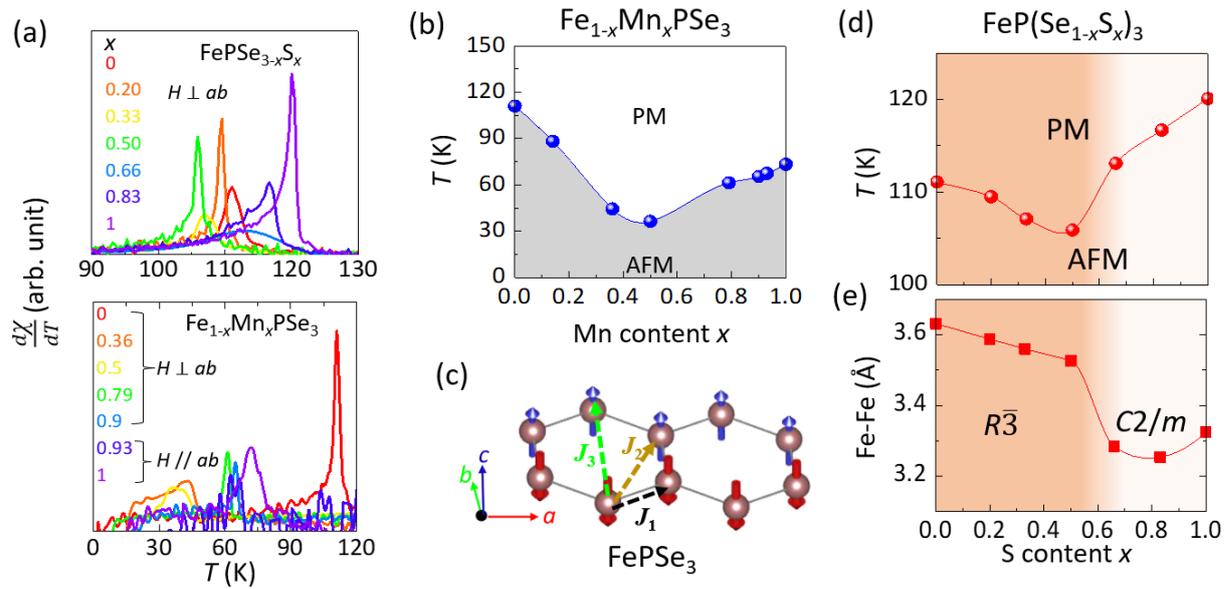

**Figure 4.** a) Temperature dependent derivative susceptibility $d\chi/dT$ for FeP(Se$_{1-x}$S$_x$)$_3$ (upper panel) and Fe$_{1-x}$Mn$_x$PSe$_3$ (lower panel) ($0 \leq x \leq 1$) samples. The peak in $d\chi/dT$ defines $T_N$. b) Doping dependence of $T_N$ for Fe$_{1-x}$Mn$_x$PSe$_3$. c) The magnetic structure of pristine FePSe$_3$ showing nearest-neighbor ($J_1$), second nearest-neighbor ($J_2$), and third nearest-neighbor ($J_3$) interactions. d) and e) Doping dependencies of d) $T_N$ and e) Fe-Fe distance for FeP(Se$_{1-x}$S$_x$)$_3$. The different colored regions in d) and e) represent different crystal structures.